\documentclass[sigconf]{acmart}

\providecommand\BibTeX{{%
 Bib\TeX}}

\usepackage[printonlyused]{acronym}
\usepackage[inline]{enumitem}
\usepackage{booktabs}
\usepackage{multirow}
\usepackage{hyperref}
\usepackage{cleveref}
\usepackage{rotating}
\usepackage{colortbl}

\acrodef{ToT}{tip-of-the-tongue}
\acrodef{KIR}{known-item retrieval}
\acrodef{LLM}{large language model}
\acrodef{ISRC}{international standard recording code}
\acrodef{QBE}{query-by-example}
\acrodef{QBV}{query-by-vocal-imitation}
\acrodef{QBH}{query-by-humming}

\crefformat{section}{\S#2#1#3} % see manual of cleveref, section 8.2.1
\crefformat{subsection}{\S#2#1#3}
\crefformat{subsubsection}{\S#2#1#3}

\newcommand{\musictot}{Music-\ac{ToT}}
\newcommand{\datasetname}{\textit{ToT$_{Music}$}}
\newcommand{\collectionname}{\textit{ToT$_{All}$}}

\newcommand{\indexwasabi}{\texttt{Wasabi}}

\newcommand{\usercreated}{\textit{User Created}}
\newcommand{\existingurl}{\textit{Extant Media}}

\newcommand{\qkeywords}{$\texttt{Keywords}$}
\newcommand{\qtitle}{$\texttt{Title}$}
\newcommand{\qtext}{$\texttt{Text}$}
\newcommand{\qtitletext}{$\texttt{Title+Text}$}
\newcommand{\reformulationN}{$\texttt{Reform}_{N}$}
\newcommand{\reformulationten}{$\texttt{Reform}_{10}$}
\newcommand{\reformulationtwentyfive}{$\texttt{Reform}_{25}$}
\newcommand{\reformulationfifty}{$\texttt{Reform}_{50}$}
\newcommand{\reformulationfull}{$\texttt{Reform}_{\infty{}}$}

\newcommand{\streamingservice}{\textit{Spotify}}
\newcommand{\partitle}[1]{\vspace{0.04in}\noindent \textbf{#1}}

\AtBeginDocument{%
  \providecommand\BibTeX{{%
    \normalfont B\kern-0.5em{\scshape i\kern-0.25em b}\kern-0.8em\TeX}}}

\setcopyright{acmcopyright}

\copyrightyear{2023}
\acmYear{2023}
\setcopyright{rightsretained}
\acmConference[SIGIR '23]{Proceedings of the 46th International ACM SIGIR Conference on Research and Development in Information Retrieval}{July 23--27, 2023}{Taipei, Taiwan}
\acmBooktitle{Proceedings of the 46th International ACM SIGIR Conference on Research and Development in Information Retrieval (SIGIR '23), July 23--27, 2023, Taipei, Taiwan}\acmDOI{10.1145/3539618.3592086}
\acmISBN{978-1-4503-9408-6/23/07}

\acmSubmissionID{123-A56-BU3}

\makeatletter
\gdef\@copyrightpermission{
 \begin{minipage}{0.3\columnwidth}
  \href{https://creativecommons.org/licenses/by/4.0/}{\includegraphics[width=0.90\textwidth]{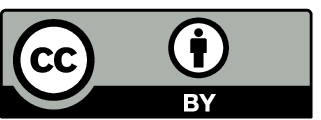}}
 \end{minipage}\hfill
 \begin{minipage}{0.7\columnwidth}
  \href{https://creativecommons.org/licenses/by/4.0/}{This work is licensed under a Creative Commons Attribution International 4.0 License.}
 \end{minipage}
 \vspace{5pt}
}
\makeatother

\begin{document}
\title{When the Music Stops: Tip-of-the-Tongue Retrieval for Music}

\author{Samarth Bhargav}
\authornote{Work done during an internship at Spotify.}
\affiliation{
  \institution{University of Amsterdam}
  \city{Amsterdam}
  \country{Netherlands}
}
\email{s.bhargav@uva.nl}
\orcid{0000-0001-5204-8514}

\author{Anne Schuth}
\affiliation{%
  \institution{Spotify}
  \city{Amsterdam}
  \country{Netherlands}
}
\email{aschuth@spotify.com}
\orcid{0000-0002-5841-2134}

\author{Claudia Hauff}
\affiliation{%
  \institution{Spotify}
  \city{Amsterdam}
  \country{Netherlands}
}
\email{claudiah@spotify.com}
\orcid{0000-0001-9879-6470}

\begin{abstract}  
  We present a study of \emph{\Acf{ToT}} retrieval for \textit{music}, where a searcher is trying to find an existing music entity, but is unable to succeed as they cannot accurately recall important identifying information. \ac{ToT} information needs are characterized by complexity, verbosity, uncertainty, and possible false memories. We make four contributions. \emph{(1)} We collect a dataset---\datasetname{}---of 2,278 information needs and ground truth answers. \emph{(2)} We introduce a schema for these information needs and show that they often involve multiple modalities encompassing several Music IR sub-tasks such as lyric search, audio-based search, audio fingerprinting, and text search. \emph{(3)} We underscore the difficulty of this task by benchmarking  a standard text retrieval approach on this dataset. \emph{(4)} We investigate the efficacy of query reformulations generated by a \acf{LLM}, and show that they are not as effective as simply employing the entire information need as a query--leaving several open questions for future research.
\end{abstract}




\maketitle

\section{Introduction}

The \emph{\Acf{ToT}} retrieval task involves identifying a previously encountered item for which a searcher was unable to recall a reliable identifier.  \Ac{ToT} information needs are characterized by verbosity, use of hedging language, and false memories, making retrieval challenging~\cite{arguello21:tot, bhargav22:tomt}.   
As a consequence, searchers resort to communities like \href{https://www.reddit.com/r/tipofmytongue/}{r/TipOfMyTongue} and \href{https://www.watzatsong.com/}{WatzatSong}, where they can post descriptions of items that they know exist but cannot find, relying on other users for help. Recent research of \ac{ToT} information needs explored how searchers pose these requests in specific domains like movies \cite{bhargav22:tomt, arguello21:tot}, or games \cite{jorgenson2020_kirgames}. \musictot{}, however, is under-explored despite being frequent: it represents 18\% of all posts made in a five-year period in the \href{https://www.reddit.com/r/tipofmytongue/}{r/TipOfMyTongue} community (cf. \cref{sec:data_collection}). Our work is motivated by the need to understand how such requests are expressed in the music domain.

 We examined the \href{https://www.reddit.com/r/tipofmytongue/}{r/TipOfMyTongue} community, focusing on requests looking for musical entities like albums, artists or songs. We show that these requests often refer to multiple modalities (cf. \cref{sec:analysis}) and thus encompass a broad set of retrieval tasks---audio fingerprinting, audio-as-a-query, lyric search, etc. In our work, we focus on song search. We create \datasetname{}\footnote{\datasetname{} (along with annotations) will be made available here: \url{https://github.com/spotify-research/tot}}: the dataset consists of 2,278 \emph{solved} information needs pertaining to a song, each of which is linked to the corresponding correct answer in the publicly available Wasabi Corpus \cite{buffa2021:wasabi}. Using \datasetname{}, we develop a schema for \musictot{} information needs to reveal what information is contained in them (cf. \cref{sec:annotation_schema}). 
In addition, we are interested in the extent to which standard text retrieval approaches are able to deal with \ac{ToT} queries. To this end, we benchmark a subset of \datasetname{} information needs\footnote{Concretely, 1.2K \textit{descriptive} information needs not containing hyperlinks -- we aimed to exclude posts where important information is not encoded in the text of the post itself.} on the Wasabi corpus, as well as Spotify search. Across both settings, the low effectiveness---compared to non-\ac{ToT} queries---of our evaluated retrieval methods underscores the necessity of novel methods to tackle this task. Lastly, we conduct a preliminary study on reformulating \musictot{} queries using GPT-3~\cite{brown20:gpt3}; we find that the task remains very challenging.

\section{Background}\label{sec:background_rw}

\partitle{\Acf{ToT} retrieval} is related to \ac{KIR} or item-re-finding~\cite{sadeghi2014}, however \ac{ToT} queries are typically issued only once---not multiple times---and importantly, lack concrete identifiers, instead relying on verbose descriptions, frequently expressed uncertainty and possible false memories~\cite{arguello21:tot, bhargav22:tomt, hagen2015_corpus, jorgenson2020_kirgames}. Approaches for simulating such queries~\cite{azzopardi_2007, elsweiler2011, kim2009} may lack realistic phenomena like false memories~\cite{hauff2012, hauff2011}, necessitating the collection of real world data. Data on a large scale is available for only one domain, movies~\cite{bhargav22:tomt}; smaller scale datasets are available for games~\cite{jorgenson2020_kirgames} and movies~\cite{arguello21:tot}. \citet{hagen2015_corpus} collect a corpus of general known-item queries, including music; however their focus was on general known-item queries and false-memories, and lacked retrieval experiments. Our focus is on the music domain, examining modalities employed by searchers and how they express \musictot{} queries. We build upon \citet{arguello21:tot} and \citet{bhargav22:tomt}, with key differences in (1) the domain---music, (2) the corpus size---millions of items instead of thousands, and, (3) reformulation experiments utilizing an \ac{LLM}. \musictot{} relates to several research areas in Music IR (MIR).

\partitle{Lyric- and text-based retrieval} involves retrieving a song using lyrics or text \cite{downie2002:theory_mir, muller2007:lyric_based_audio_retrieval}. Techniques to handle \textit{misheard} lyrics are common \cite{ring2009:misheard_lyrics, xu2012:lyric_search_mishearing, xu2009:lyric_phonetic, ye2020:cross_lan_misheard}, including modeling speech sounds~\cite{hirjee2010:misheard_probalistic}, which may be insufficient, since \ac{ToT} queries can contain \textit{descriptions} of lyrics, requiring semantic methods \cite{sasaki2014:lyrics_latent}, or utilizing the audio itself \cite{yu2019:cross_model_audiolyrics}.  Apart from lyrics, \musictot{} queries are frequently free-form natural language queries (cf. \cref{sec:analysis}), requiring methods that can retrieve audio using text, as well as tags, genre or human-generated descriptions \cite{oncescu2021:audretr_nl_queries, koepke2022:audretr_benchmark, elizalde2019:crmodal_joint_audio, zhu2022:content_based_framework, doh2022:universal_text_to_music}. 

\partitle{Content-based audio retrieval}~\cite{foote1997:content_based_audio} includes \ac{QBE} \cite{helen2007:qbe}, where the audio is being queried as-is, e.g. audio fingerprinting \cite{haitsma2002:audfinger_seminal}. Alternatively, users can \textit{imitate} the wanted audio by vocalizing it, termed \ac{QBV} \cite{zhang2018:qbv, kim2019:vocalimitation}, which includes \ac{QBH} \cite{ghias1995:qbh}. \ac{ToT} queries frequently contain references to user created audio-clips as well as existing media like audio contained in videos (cf. \cref{sec:analysis}). 

\partitle{Other modalities} like videos may need to be handled as well, necessitating multi-modal or cross-modal (retrieving one modality using another) methods \cite{simonetta2019:multimodal_mir}, e.g. retrieving audio using video \cite{wang2016:cross_modal_survey, hong2017:cross_modal_audiovideo}. 
Approaches to solve \musictot{} have to account for multiple modalities and free-form natural language including noise, e.g., uncertainty \cite{arguello21:tot} and/or false memories \cite{arguello21:tot, jorgenson2020_kirgames}.

\section{Methodology}\label{sec:methodology}

\subsection{Data Collection}\label{sec:data_collection}

\partitle{Gathering \collectionname{}.}\label{sec:data_collection_full} We gathered posts made across 2017-2021 in the \href{https://www.reddit.com/r/tipofmytongue/}{r/TipOfMyTongue} community, yielding 503,770 posts (after filtering out posts not marked \textit{Solved} or \textit{Open}), each containing two fields: \textit{title} and \textit{description}. We extracted text categories from the title, e.g. \texttt{SONG} from \emph{"[SONG] Slow dance song about the moon?"}. We manually identified a set of 11 overarching \emph{music-focused} categories (e.g. \textit{Music Video}, \textit{Band}, \textit{Rap Music}). We discarded the remaining non-music posts, resulting in \collectionname{}: 94,363 (60,870 solved and 33,493 unsolved) \musictot{} posts. These posts form a large proportion---18.73\%---of the 503K posts we started out with. 

\partitle{Extracting \datasetname{}.}\label{sec:data_collection_with_ans}  We extracted answers from \textit{Solved} posts following \citet{bhargav22:tomt}, retaining \textit{Solved} posts which have a URL as an answer. If the URL points to a \textit{track} on \streamingservice{}, obtaining the answer was trivial. Otherwise, the \texttt{title} portion of the markdown inline URLs, formatted as [\texttt{title}](\texttt{url}) (with title often formatted as `Artist-Song') was used as a query to the \streamingservice{} search API. Since the API returns multiple results, we created a classifier\footnote{Random Forest classifier, parameters selected with grid search on \{10, 20, 30, 40, 50\} estimators, max depth \{2, 3, 4\} and min/max scaled features.} with 31 features based on the scores of the retriever, the edit distances between \texttt{title} and artist name, song title, etc. We used the classifier to predict if a \texttt{title} matches the track and artist, scoring 100\% on precision on a held out set of 100 samples. Low-confidence candidates were filtered out. This left us with a set of 4,342 posts with \streamingservice{} tracks as answers. Lastly, we only retained those posts where the \acs{ISRC}\footnote{The \ac{ISRC} is a standardized code for uniquely identifying recordings.} of the answer track is also present in the Wasabi Corpus~\cite{buffa2021:wasabi}: a total of 2,278 posts. We call this collection \datasetname.

\partitle{Gathering reformulations.}\label{sec:data_reformulations} We gathered reformulations for all posts in \datasetname{} by prompting GPT-3~\cite{brown20:gpt3}\footnote{Model: \texttt{text-davinci-003}, with temperature 0.7} with the respective post description and a word count limit: \emph{\texttt{<description>} Summarize the query above to \texttt{<N>} words, focusing on musical elements}. We used $N=\{10, 25, 50\}.$\footnote{Based on manual inspection, we discarded $N=5$ (too few words for a cohesive query, leading to crucial information being left out) and $N=100$ (model hallucinations).} We also employed a prompt without a specific word limit: \emph{\texttt{<post description>} Shorten the query above, focusing on musical elements.} 

\subsection{Music-ToT Schema}\label{sec:annotation_schema}

Our annotation process involved three steps. We first developed and then refined a schema to describe \musictot{} information needs; in the final step, we annotated 100 samples from \datasetname{}.

\partitle{Developing the schema in 2 steps.} 
A preliminary study conducted with one author (self-rated music expertise 7 out of 10) and two volunteers (music expertise 8/10 and 7/10 respectively) involved assigning one or more labels to 78 sentences from 25 randomly sampled posts from \datasetname{}. We focused on developing new labels specific to \musictot{}, while also re-using labels from \citet{arguello21:tot}: specifically the \textit{Context} labels, pertaining to the context an item was encountered in   (\emph{Temporal Context}, \emph{Physical Medium}, \emph{Cross Media}, \emph{Contextual Witness}, \emph{Physical Location}, \emph{Concurrent Events}), and \textit{Other} annotations (\emph{Previous Search}, \emph{Social}, \emph{Opinion}, \emph{Emotion}, \emph{Relative Comparison}). The latter are generally applicable across \ac{ToT} information needs. This preliminary study revealed 25 new music labels, in addition to 11 labels from prior work ($6\times{}$\textit{Context} and $5\times{}$\textit{Other}).  In the second step, the three authors (self-rated musical expertise 7, 6 and 5 respectively) of this paper labeled 110 sentences (20 posts from \datasetname{}) to validate the schema. Based on our results and discussions, we combined a few finer-grained categories with low support into more general categories, e.g. specific musical elements like \emph{Rhythm} / \emph{Repetition}, \emph{Melody}, \emph{Tempo}, etc., were combined to \emph{Composition}, resulting in \textbf{28 labels in total}.

\partitle{Annotating.} Lastly, in step 3, two authors employed the final schema to annotate $536$ sentences corresponding to $100$ posts. The resulting labels, their frequency, category, inter-rater agreement (Cohen's $\kappa$ \cite{cohen1960:kappa, artstein2008:kappa}) along with their description and an example, are presented in Table \ref{tab:schema}.

\begin{table*}[ht!]
\footnotesize
\centering
\caption{Annotation Schema: Label, frequency of occurrence in 100 submissions / 536 sentences (\textbf{F}), annotator agreement ($\kappa$) and description of label, along with an example for each label.}
\label{tab:schema}

\begin{tabular}{@{}llrrp{6.5cm}p{6cm}@{}}
\toprule
& \textbf{Label}                   & \textbf{F}   & $\mathbf{\kappa}$ & \textbf{Description} & \textbf{Example}                                                                                                                                                                                                                                   \\ \midrule
\multirow{30}{*}{\begin{sideways}\textbf{MUSIC ANNOTATIONS}\end{sideways}}            & \cellcolor{white} Composition             & 87  & 0.74     & Describes (part of) the composition of a piece of music including rhythm, melody, tempo, pitch, chords, notes, and keys; or how they are composed into a cohesive piece of music.  & \ldots playing the same \textit{major-key pattern over each chord in a fairly simple repeating loop.}  \\
                                                                                         &\cellcolor{gray!22} Genre                   & \cellcolor{gray!22} 77  & \cellcolor{gray!22} 0.92     & \cellcolor{gray!22} References a genre.  & \cellcolor{gray!22} It sounded like a \textit{reggae/ska} type beat \\
                                                                                          & \cellcolor{white} Music Video Description & 75  & 0.89     & Describes a music video associated with a song. &  \textit{However, once the music starts, the store is lit up and the tone shifts completely as everything in that store has a pastel colour scheme. }               \\
                                                                                          & \cellcolor{gray!22} Lyric Quote             & \cellcolor{gray!22}65  & \cellcolor{gray!22}0.89     & \cellcolor{gray!22}Directly quotes lyrics that the user overheard, not including sounds / vocalizations &\cellcolor{gray!22}  \ldots it wasn't until he said something about the \textit{``just somebody that I used to know''} song that I \ldots                                                                                                                                                            \\
                                                                                          & \cellcolor{white} Story/Lyric Description & 60  & 0.71     & Describes either the story conveyed by the lyrics, or the gist of the lyrics instead of directly quoting it. & The song is \textit{a woman singing to/about a man that she was in love with and died, I think he was in the military and got killed and she had a baby at home}? \\
                                                                                          & \cellcolor{gray!22} Artist Description      & \cellcolor{gray!22}54  & \cellcolor{gray!22}0.92     &\cellcolor{gray!22} Describes the artist.  & \cellcolor{gray!22}\textit{He was maybe a tad overweight, shaggy hair, maybe curly.} \\
                                                                                          & \cellcolor{white} Time Period / Recency   & 49  & 0.89     & References the time period the user thought the music was produced.  & \textit{Late 90s-early 2000s} hip hop song that sounds similar to clip   \\
                                                                                          & \cellcolor{gray!22} Instrument              & \cellcolor{gray!22}30  & \cellcolor{gray!22}0.86     & \cellcolor{gray!22} Mentions instruments that were overheard.     &  \cellcolor{gray!22} The guy performing was at a \textit{keyboard/piano} \ldots \\
                                                                                          & \cellcolor{white} Vocals                  & 28  & 0.69     & Describes the voice or vocal type.  & \textit{High pitched but kind of floaty female vocals}, a bit \ldots \\
                                                                                          & \cellcolor{gray!22} Name                    & \cellcolor{gray!22} 23  & \cellcolor{gray!22} 0.81     & \cellcolor{gray!22} Describes a song/artist/album name, what it resembles/contains, or what the searcher remembers of it.       &  \cellcolor{gray!22} \ldots \textit{the name of the song was brief, one nordic word}.                \\
                                                                                          & \cellcolor{white} Popularity                    &  18	 & 0.83     & Describes the popularity of the music, artist, album or music video.  & I'm surprised I can't find it since I can remember many specific lyrics, I guess it's \textit{more obscure} \\
                                                                                          & \cellcolor{gray!22} Recording                    &   \cellcolor{gray!22} 15 & \cellcolor{gray!22} 0.80	     &\cellcolor{gray!22}   A description or reference to user-created content & \cellcolor{gray!22} \textit{I did a vocaroo of the tune, sorry about my voice and any possible background guinea pig noises:} \texttt{URL}\\
                                                                                          & \cellcolor{white} Language / Region     &  14 & 0.92     &  Either mentions the language of the piece of music and/or references a particular region like state, country, etc. &  A \textit{Japanese} song that I don't remember any words to or how the tune goes at all, \\
                                                                                          & \cellcolor{gray!22} Album Cover     &  \cellcolor{gray!22} 5 & \cellcolor{gray!22} 1.00     & \cellcolor{gray!22} Describes the album cover.            & \cellcolor{gray!22} \textit{On the cover there was also a cyan teal line going along the bottom with white text in it.}        \\
                                                                                          & \cellcolor{white} Song Quality / Type                     &  4 & 0.00     & Describes the type of music (original/cover, live/recorded) or the production quality (professional, amateur, etc.)   & \textit{Live Cover} of All Along the Watchtower where \ldots \\
                                                                                          \midrule
\multirow{24}{*}{\begin{sideways}\textbf{CONTEXT et al. ANNOTATIONS}\end{sideways}}
 & \cellcolor{gray!22} Uncertainty             & \cellcolor{gray!22} 162 & \cellcolor{gray!22} 0.79     &\cellcolor{gray!22}  Conveys uncertainty about information described. & \cellcolor{gray!22} \textit{I don't know} what genre the song was, it was fairly calming and \textit{I feel like} it couldve been on tiktok but \textit{I don't really know}.  \\
 & \cellcolor{white} Social                  & 54  & 0.77     & Communicates a social nicety. &   \textit{Any help appreciated!} \\
                                                                                          & \cellcolor{gray!22} Opinion          &  \cellcolor{gray!22} 43	 & \cellcolor{gray!22} 0.44     &  \cellcolor{gray!22}  Conveys an opinion or judgment about some aspect of the music. &  \cellcolor{gray!22} I don't remember the lyrics or title, only that \textit{it was a kind of angsty teen ``I want to set the world on fire''} \\
                                                                                          & \cellcolor{white} Temporal Context        & 36  & 0.87     & Describes when the music was heard, either in absolute terms or relative terms.  & \ldots I heard like in a billion videos \textit{6 years ago}. \\
                                                                                          & \cellcolor{gray!22} Listening Medium        & \cellcolor{gray!22} 26  & \cellcolor{gray!22} 0.75     & \cellcolor{gray!22} References the medium associated with the item. (e.g., radio, streaming service, etc) & \cellcolor{gray!22} I heard it on the \textit{radio} a couple of times in  \ldots\\
                                                                                          & \cellcolor{white} Embedded Music        & 26  & 0.58     & References or describes extant media (e.g., Youtube / Twitch URL), including timestamps. & I do have a \textit{video with the song (this video at around minute 4:21: \texttt{URL})} \\
                                                                                          		
                                                                                          & \cellcolor{gray!22} Other Cross Media        & \cellcolor{gray!22} 26  & \cellcolor{gray!22} 0.19     & \cellcolor{gray!22} Describes exposure to the piece of music through different media, excluding  other Cross Modal labels  & \cellcolor{gray!22} \ldots I'm pretty sure was performed on one of the \textit{early seasons of Glee or maybe Smash}. \\ 
                                                                                          & \cellcolor{white} Previous Search         & 25  & 0.67     & Describes a previous attempt to find the item, including negative results (i.e., it is not song X).    & \textit{I've tried humming it into shazam and other sites, looking up the two generic lyrics I remember, even doing those rhythm tapping things and nada}  \\
                                                                                          & \cellcolor{gray!22} Relative Comparison     & \cellcolor{gray!22} 25  & \cellcolor{gray!22} 0.77     & \cellcolor{gray!22} Describes a characteristic of the music in relative (vs. absolute) terms, by explicitly comparing it with another song / artist / album. & \cellcolor{gray!22} The melody I remember \textit{resembles the beginning of the song "Run to the hills" by Metallica} \\
                                                                                          & \cellcolor{white} Emotion     &  25 & 0.05     & Conveys or describes how a piece of music made the viewer feel & Even talking about it \textit{makes me tear up.} \\
                                                                                          & \cellcolor{gray!22} Concurrent Events     & \cellcolor{gray!22} 18  & \cellcolor{gray!22} 0.09     & \cellcolor{gray!22}  Describes events relevant to the time period when music was encountered, but excluding descriptions of the music itself. & \cellcolor{gray!22} \ldots when \textit{I was driving down the country} but for the life of me can’t remember the name. \\
                                                                                          & \cellcolor{white} Physical Location     &  9 & 0.61     & Describes physical location where music was encountered. & \ldots record a 9 second portion of this song at \textit{a Marriott hotel bar in downtown Chicago}  \ldots \\
                                                                                          & \cellcolor{gray!22} Contextual Witness     &  \cellcolor{gray!22} 9 & \cellcolor{gray!22} 0.49     & \cellcolor{gray!22} Describes other people involved in the listening experience. & \cellcolor{gray!22} A few years back, \textit{a friend of mine} showed me an \ldots         \\
                                                                                          \bottomrule
\end{tabular}%
\end{table*}

\section{Data Analysis}\label{sec:data_description}\label{sec:analysis}

We now first discuss Table \ref{tab:schema}, followed by a brief discussion about the modalities present in the whole collection, \collectionname{}.

\partitle{Annotation results.} 
Among the music-focused annotations, \textit{Genre} and \textit{Composition}, a description of musical elements and how they fit together, are the two most frequent labels. This is followed by \textit{Music Video Description}, and either direct quotes (\textit{Lyric Quote}) or a description of the lyrics (\textit{Story/Lyric Description}) further highlighting the different information needs that need to be addressed i.e., lyric search, text search and multi-modal search. However, a simple extraction of \textit{Genre} and metadata such as \textit{Time Period/Recency}, \textit{Instrument}, etc., may not be useful without considering the most frequent label, \textit{Uncertainty}. Search systems therefore would have to handle these elements, as well as consider potential false memories. Furthermore, annotations like \textit{Social}, \textit{Opinion} are also fairly common occurrences in our data, which may have limited utility for retrieval~\cite{arguello21:tot}, motivating reformulations (cf. \cref{sec:data_reformulations}). Searchers also express their queries in terms of other music entities in a \textit{Relative Comparison}, and describe \textit{Previous Search} attempts, explicitly ruling out certain candidates. References to other modalities like user created clips (\textit{Recording}) or existing media (\textit{Embedded Music}) also pose a challenge. We now explore this challenge with a brief study of references to external content in the entire collection, \collectionname{}.

\partitle{Cross-modal references} \musictot{}, like other \ac{ToT} domains, contains cross-modal and media references \cite{arguello21:tot}, where a searcher refers to external content. We here show that \musictot{} posts in particular contain such references frequently. To this end, we gathered frequent websites that appear in \collectionname{}. One author manually labeled these as one of: \begin{enumerate*}
    \item \usercreated: a clip uploaded by a user, e.g., \href{https://vocaroo.com/}{Vocaroo}, \href{https://clyp.it/}{Clyp.it}, \href{https://www.drive.google.com/}{Google Drive}, \href{https://www.dropbox.com/}{Dropbox}, \href{https://instaud.io/}{Instaudio}, \href{https://musiclab.chromeexperiments.com/}{musiclab}, \href{https://onlinesequencer.net/}{Onlinesequencer}, \href{https://streamable.com/}{Streamable}, \href{https://www.speakpipe.com/}{Speakpipe}.
    \item \existingurl: a clip unlikely to be uploaded by a user, e.g. an existing clip, corresponding to content/social media websites like Spotify, Twitch, Tiktok, or YouTube.
    \item \emph{Other URL}:  Not belonging to the previous two categories. 
\end{enumerate*} We find that \existingurl~ forms a larger proportion of queries (19K, 20.9\%) compared to \usercreated~ queries (14K, 15.3\%), with a small number of posts containing references to both types (1.1\%). Therefore, \musictot{} information needs are inherently multi-modal. We characterize the remaining 57.7\% of queries as \textit{descriptive} queries, which include references to lyrics, or story descriptions (cf. \cref{sec:annotation_schema}).  In summary, \musictot{} information needs are characterized by uncertainty and multi-modality, requiring methods like text-based audio retrieval, content based audio retrieval/fingerprinting and multi- or cross-modal retrieval.

\section{Benchmarks}

\subsection{Experimental Setup}\label{sec:exp_setup}

\partitle{Corpora.}\label{sec:corpus} We run experiments on two corpora. The first is the \indexwasabi{} 2.0 Corpus \cite{buffa2021:wasabi, buffa2021:zenodowasabi}. It consists of 2M commercial songs from 77K artists and 200K albums. Crucially, \begin{enumerate*}
    \item songs have the \ac{ISRC} linked, enabling linking to data in \streamingservice{};
    \item it is an open dataset, consisting of rich information that includes lyrics, extensive metadata, and music snippets 
\end{enumerate*}. We index the \textit{Song Name}, \textit{Artist Name} and \textit{Lyrics}\footnote{We also experimented with other fields like \textit{Album Title}, but saw no improvement in retrieval effectiveness.} of all songs using Elasticsearch (BM25 with default parameters). 
The second corpus corresponds to the \streamingservice{} US catalog, consisting of hundreds of millions of tracks. The \streamingservice{} search system \cite{neural_instant_search:hashemi2021} utilizes multiple retrieval stages (including lexical- and semantic search) and incorporates historic log data for retrieval purposes. %We term this system \internalindex.

\partitle{Queries.}\label{sec:queries} We conducted experiments on the 1,256 posts (849 train, 191 validation, and 216 test) from \datasetname{} that contain no URLs in the post title or post text; we make this choice as in the most extreme case, the entire post may contain just a URL, requiring audio-based search while we focus on text-based methods. From each post, we create different \emph{queries} and label them as follows: \begin{enumerate*}
    \item \texttt{Title}: using the post title only;
    \item \texttt{Text}: post text;
    \item \texttt{Title+Text}: title \& text concatenated; and finally,
    \item \qkeywords{}: extracting up to ten keywords from the post text\footnote{Keywords were deduplicated with threshold = $0.2$ and algorithm =\texttt{seqm}.} with Yake~\cite{campos2018:yake_ecir}; 
    \item \reformulationN{}: reformulations with $N=\{10,25,50,\infty{}\}$. 
\end{enumerate*}

\partitle{Evaluation.} We report Recall@K, equivalent to Success@K (i.e., one correct answer) for $K=\{10,100,1000\}$ on \indexwasabi{}. 
All reported results are on the test set. For \streamingservice{} search we describe the observed trends (due to the proprietary nature of the system).

\begin{table}[!htb]
\centering
\caption{Overview of retrieval experiments on \indexwasabi{}, using Elasticsearch (BM25).}
\label{tab:baseline}
% \resizebox{0.4\textwidth}{!}{%
\begin{tabular}{@{}rccc@{}}
\toprule
\textbf{Query} & \textbf{S@10} & \textbf{S@100} & \textbf{S@1000} \\
\midrule
   \qtitle                  & 0.0370                & 0.0833                 & 0.1389                  \\
                                 \qkeywords              & 0.0231                & 0.0463                 & 0.0787                  \\
                                 \qtext                   & 0.0139                & 0.0648                 & 0.0926                  \\
                                 \qtitletext             & 0.0324                & 0.0833                 & 0.1713                  \\
                                 \midrule % \cmidrule{2-5}
                                 \reformulationten        & 0.0139 & 0.0509 & 0.1204 \\
                                 \reformulationtwentyfive & 0.0278 & 0.0602 & 0.1389 \\
                                 \reformulationfifty      & 0.0185 & 0.0741 & 0.1389 \\
                                 \reformulationfull       & 0.0139 & 0.0741 & 0.1574 \\ 
\bottomrule
\end{tabular}%
% }
\vspace{-0.3in}
\end{table}

\subsection{Results}

Table~\ref{tab:baseline} provides an overview of our \indexwasabi{} results. 

\partitle{Post parts as query.} The low success across queries and $K$ underscores the difficulty of the task. On \indexwasabi{}, \qtitle{} queries are more effective than \qtext{} queries---increased verbosity leads to retrieval failure. However, the text may indeed contain data useful in retrieval, with comparable or higher effectiveness scores for \qtitletext{} over \qtitle{} at $K=\{100, 1000\}$, motivating keyword extraction: crucial details might be present in the text, but including the entire need as a query might harm effectiveness. Our keyword selection method though fails to outperform other queries except for \qtext{} on S@10. 

On \streamingservice{} search we observe a different trend: \qtitletext{} is the most effective query followed by \qtitle{}. 

\partitle{LLM reformulations as query.}\label{sec:res_reform} Examining Table \ref{tab:baseline}, reformulations have limited success compared to \qtitle{} queries. \reformulationtwentyfive{} and \reformulationfifty{} perform as well as \qtitle{} on S@1000, with \reformulationfull{} outperforming it. While \qkeywords{} beat all but \reformulationtwentyfive{} on S@10, it is outperformed by reformulations on S@100 and S@1000. On \streamingservice{} search, we find that reformulations fare worse than \qtitle{} queries for S@10, but see limited success on S@100, with \reformulationtwentyfive{} and \reformulationfifty{} achieving higher effectiveness. Most importantly, there is no ideal $N$ on either index, with varying success across metrics. We thus conclude that in our study, reformulations generated using state-of-the-art \ac{LLM}s have only mixed success.

\section{Conclusions}

We explored Tip-of-the-Tongue retrieval for music. 
Of the 94K posts corresponding to \musictot{} information needs from an online community for \ac{ToT} requests, we linked 2,278 posts to the corresponding answers in the Wasabi corpus, resulting in \datasetname{}, thus enabling further research for this challenging task. 

We iteratively developed and refined a \musictot{} schema that contains 28 fine-grained labels as shown in Table~\ref{tab:schema}. Labeling 100 posts using this schema, we showed that users express uncertainty frequently, and almost as often refer to other modalities. 
We benchmarked a subset of 1.2K \textit{descriptive} queries from \datasetname{}, and highlight the difficulty of the task. Future work should leverage cross- and multi-modal retrieval as well as better approaches for reformulations.

\begin{acks}
The authors would like to thank Gulfaraz Rahman and Ruben van Heusden for helping with the preliminary annotation work. The authors also thank Daniel Lazarovski and Humberto Corona Pampín for their input. Part of this research was supported by the NWO Innovational Research Incentives Scheme Vidi (016.Vidi.189.039). All content represents the opinion of the authors, which is not necessarily shared or endorsed by their respective employers and/or sponsors. 
\end{acks}

\bibliographystyle{ACM-Reference-Format}
\balance
\bibliography{references}

\end{document}